# Manipulation, trapping, splitting and merging of water and aqueous bio-droplets by photovoltaic optoelectronic tweezers

Andrés Puerto,[1] Angel Méndez,[2] Luis Arizmendi,[1] Angel García-Cabañes,[1*] and Mercedes Carrascosa[1*]

[1]*Departamento de Física de Materiales, Facultad de Ciencias, Universidad Autónoma de Madrid, Calle Francisco Tomás y Valiente 7, E-28049 Madrid, Spain*
[2]*Departamento de Mecánica de Fluidos y Propulsión Aeroespacial, ETSIAE, Universidad Politécnica de Madrid, Plaza Cardenal Cisneros 3, E-28040 Madrid, Spain*

[*]E-mails: m.carrascosa@uam.es
angel.garcia@uam.es

## Abstract

Optical and optoelectronic techniques for micro- and nano-object manipulation are becoming essential tools in nano- and bio-technology. A remarkable optoelectronic technique that has experimented a strong development in the last few years is the so called photovoltaic optoelectronic tweezers. It is based on the light-induced electric fields generated by the bulk photovoltaic effect in certain ferroelectrics such as $LiNbO_3$. The technique is simple and versatile, enabling a successful manipulation of a large variety of micro- and nano-objects with only optical control, without the need of electrodes or power supplies. However, it is still a challenge for this tool, to handle objects immersed in aqueous solution due to the electric screening effects of polar liquids. This has hindered their application in biotechnology and biomedicine where most processes develop in aqueous solution. In this work, a new efficient route to overcome this problem has been proposed and demonstrated. It uses photovoltaic optoelectronic tweezers to manipulate aqueous droplets, immersed in a non-polar oil liquid, but hanging at the interface air-oil. In this singular configuration, the high electric fields generated in the photovoltaic substrate allow a simple and flexible manipulation of aqueous droplets controlled by the light. Droplet guiding, trapping, merging and splitting have been achieved and efficient operation with water and a variety of bio-droplets (DNA, sperm, and PBS solutions) have been demonstrated. The reported results overcome a main limitation of these tweezers to handle bio-materials and promises a high potential for biotechnological and biochemistry applications including their implementation in optofluidic devices.

# 1. Introduction

Optical manipulation has become a key tool for nano- and bio-technology. Main optical techniques are optical [1, 2] and optoelectronic tweezers [3, 4]. Optical tweezers are very efficient in the manipulation of single dielectric particles such as silica or polystyrene beads [5, 6] but, they use high intensities that can damage bio-systems [7, 8]. In turn, optoelectronic tweezers, that use a photoconductor substrate under an AC electric field, operate at lower intensities providing larger forces and are better adapted for massive particle manipulation. However, they need electrodes and power supplies that complicate the devices. In both approaches, the trapping effects disappear when the light is turned off preventing their use to fabricate stable traps or patterns. Recently, a new optoelectronic approach, that uses a patterned photoconductive coating to overcome this latter problem, has been proposed demonstrating remarkable results although the substrate fabrication is considerably more complicated [9].

In the last decade another optoelectronic technique, the so called photovoltaic optoelectronic tweezers (PVOT), has been introduced [10–13]. They are simpler because operate at low light intensity without electrodes or power supplies. Recently, PVOT have experimented a remarkable development, succeeding in the manipulation, trapping and patterning of a large variety of micro- and nano-objects including particles (metal, dielectric, organic…) (see References [14, 15] and references therein) and nonpolar liquid (oil) droplets [16–18]. They are based on the bulk photovoltaic effect that allows the light-induced generation of very high electric fields patterns (100-200 kV/cm) inside certain ferroelectrics such as $LiNbO_3$:Fe (LN:Fe) [19]. These photo-induced electric fields extend outside the active optical material and are able to manipulate the micro- and nano-objects, mainly through dielectrophoretic forces [14, 20].

The technique is very efficient for massive manipulation and patterning of nano-objects immersed in a non-polar fluid such as air, tetradecane or hexane [12, 14]. Moreover, at difference with other optical/optoelectronic techniques, the photovoltaic (PV) field remains after illumination giving stability to the patterns. The PVOT technique also shows remarkable reconfiguration capabilities [21, 22]. These properties allows efficient fabrication of a variety of nanoparticle patterns for different applications such as photonic devices [23] or plasmonic platforms [24]. However, PVOT find serious difficulties to handle objects inside polar liquids such as aqueous solutions, due to their electric field screening effects, what seriously limit the potential applications in bio-photonics and biotechnology. Anyhow, a few interesting applications have been already reported, involving cells [25, 26], bacteria [27] and pollen and spores grains [28]. Then, a new route to apply extensively the remarkable capabilities of those optoelectronic tweezers in aqueous solutions would be a key advance for biotechnological functionalities and would extend still more the applications of this powerful technique.

The novel strategy proposed in this paper consists in manipulating droplets of bio-aqueous solutions inside a nonpolar immiscible oil. Since recent trends in microfluidics use droplets to miniaturize the devices (droplet or digital microfluidics), our proposal open the door to its application in this vast and very active field [29, 30] and, in particular, to its application for lab-on-a-chip devices overwhelmingly used in biotechnology [31–34]. In fact, an important issue for these technologies are the development of efficient methods to manipulate the droplets inside the inactive/passive fluid medium. Among the diverse approaches [35], optoelectronic methods, such as that proposed here, have been considered an efficient alternative.

Hence, in this work we have proposed and investigated the use of photovoltaic optoelectronic tweezers for manipulating water and aqueous bio-droplets inside a non-polar liquid medium. There has been only one previous interesting work that has attempted to handle water droplets in air [16] that could be considered a non-polar fluid. However, the experiments have failed unless the substrate is covered with a hydrophobic coating [36] because the droplets in contact with the substrate increase their wettability and become immobile. In the present approach, this problem is overcome because, during manipulation, the droplets have no contact with the substrate. Two configurations of the optoelectronic substrate, the so called parallel and perpendicular configurations [14], have been tested. Experiments of droplet guiding, splitting, and trapping, either at the interface or on the substrate are reported. Moreover, merging of two different droplets has been also achieved. The experiments have been developed for deionized water and for aqueous bio-droplets containing DNA and sperm solutions. A discussion on the physics of the manipulation process and on the possibilities of the technique is also included.

## 2. Physical basis of photovoltaic optoelectronic tweezers

PVOT are based on the bulk photovoltaic effect (BPE) presented by certain ferroelectric crystals (mainly lithium niobate doped with iron [14] but also with cupper [37] impurities). It consists in the appearance of an electric charge transport in the direction of the polar ferroelectric axis (*c*-axis) when the crystal is illuminated with visible light. It is due to asymmetric electron photoexcitation from donor (usually $Fe^{2+}$) impurities to the conduction band giving rise to a photovoltaic electric current density along the polar *c*-axis that can be written as [14]:

$$j_{pv} = e\alpha\phi l_{pv} \quad (1)$$

where, $\phi = I/h\nu$, is the photon flux, $\alpha$ the absorption coefficient and $l_{pv}$ the photovoltaic effective drift length. Photo-excited electrons finally get trapped in acceptor impurities (usually [$Fe^{3+}$]). This charge transport produces a light-induced electric field $E_{pv}$ that, in steady state, writes:

$$E_{pv} = \frac{j_{pv}}{en\mu} \quad (2)$$

*n* being the steady-state density of photo-excited electrons, $\mu$ the electron mobility and $j_{pv}$ the current density given by Equation 1. This field extends near the surface outside the crystal as an evanescent field that can act either on neutral and charged micro- or nano-objects through dielectrophoretic (DEP) or electrophoretic (EP) forces, respectively, as explain in detail in previous works [14, 15]. There are two main useful geometries for the photovoltaic substrates, the parallel (*x*- or *y*-cut crystals) and perpendicular (*z*-cut crystals) configurations, which have the polar axis parallel or normal to the active surface, respectively [38]. The structure of the evanescent electric field is different for these two geometries (see Reference [39] and simulations of section 5) and so it is important to choose the adequate substrate orientation depending on the kind of the droplet manipulation process pursued.

## 3. Experimental strategy for aqueous droplet manipulation

As already mentioned, we propose a strategy for the PVOT manipulation of biological material, by acting on aqueous bio-droplets as highly polarizable "microparticles", immersed in a nonpolar immiscible oil, thus avoiding direct screening of the photo-induced electric fields. Fig. 1 shows the experimental arrangement used, which facilitates explaining the novel strategy. In Fig. 1(a), a scheme of the optical system structure to operate with PVOT has been drawn. Basically, it consists of an optical microscope in which a laser beam is coupled through the objective lens of the microscope so that it can be focused on the active substrate. Specifically, we have used 1 mm thick LN:Fe plates with 0.25% mol Fe concentration. Two CMOS cameras recorded a top and a lateral view of the sample holder, respectively. This sample holder [inset in Fig. 1(a)] is a glass cuvette with optical quality surfaces containing a suitable transparent non-polar liquid (paraffin oil with density $\rho \approx 0.876$ g/cm$^3$ and dielectric constant $\varepsilon$=4.6-4.8, in our case). The LN:Fe plate is placed at the bottom of the cuvette and the aqueous droplets are added with a micropipette.

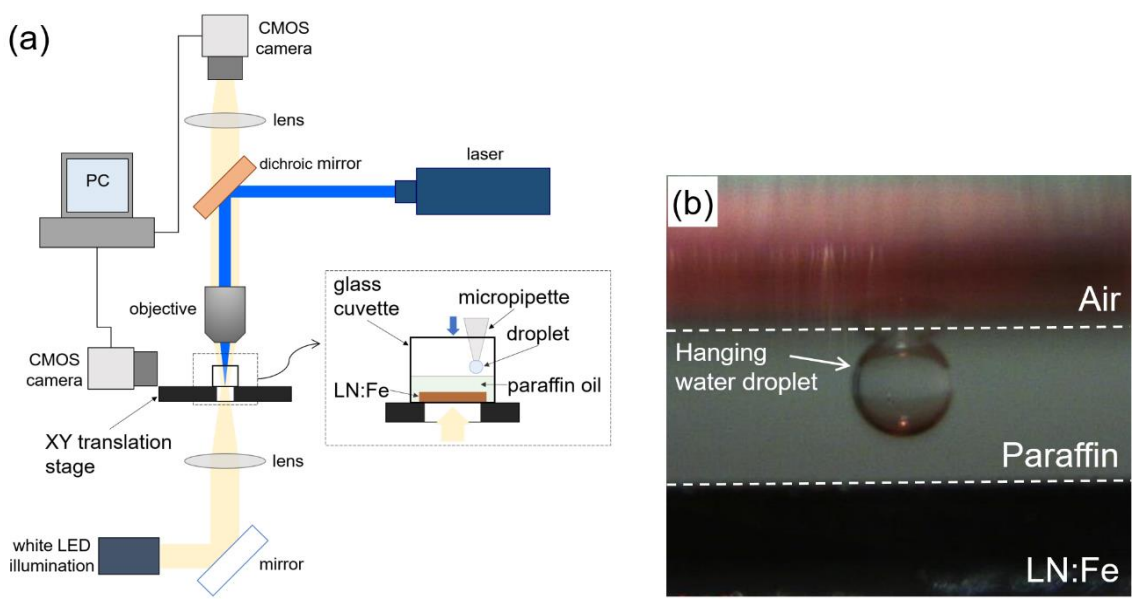

FIG. 1: (a) Schematics of the optical microscope system of the PVOT used, showing in detail the sample holder as an inset. (b) Photograph of the lateral view of the glass cuvette containing a hanging water droplet. The boundaries between the different media have been indicated with discontinuous white lines.

Since water density is higher than paraffin oil density, a water droplet immersed inside the oil would fall onto the substrate by the gravity action. However, in our strategy the aqueous droplet is deposited on the oil surface, and stays hanging from the paraffin-air boundary, as shown in Fig. 1(b), by the compensation of the apparent weight downward force with a net upward force. This latter force results from the sum of the air-water, water-paraffin and air-paraffin surface tension forces as it is explain in detail in Reference [40]. This fact is a key advantage of our method because it avoids droplet immobilizing on the substrate, due to the high wettability of water on LN even enhanced by the light-induced electric field [36]. This way easy and flexible optical manipulation of water droplets by a single laser beam will be achieved. The manipulation capabilities extend to droplets of a variety of aqueous liquids of interest in bio-photonics technology such as deionized water, PBS (cell culture medium), and aqueous solutions containing sperm, or even fluorescein tagged DNA.

### 4. Droplet manipulation results.

As mentioned, manipulation experiments pursue different kind of processes such as controlled droplet moving, trapping, merging and splitting. The two substrate geometries (*z*- and *x*-cut) for which the symmetry of the photo-induced electric fields is different have been tested finding the best configuration for each manipulation process.

### 4.1 Droplet moving and trapping on the PV substrate

The first set of manipulation experiments has been developed with droplets from a variety of available aqueous liquids (water and DNA, sperm and PBS aqueous solutions) and using the perpendicular configuration (*z*-cut crystals). The droplet volumes ranged between 0.1 and 1 µL. A diode Gaussian laser beam ($\lambda$ = 488nm) is focused on the substrate to a diameter of a 160 µm ($4\sigma$) and with a light intensity of $2.7 \times 10^6$ Wm$^{-2}$. An illustrative experiment with a droplet of an aqueous solution of fluorescein tagged DNA is shown in Fig. 2, where a lateral view of the process is presented by a sequence of images. Initially, still in the dark, the droplet has been pipetted and stays hanging from the paraffin-air interface [Fig. 2(a)]. Then, the laser beam is focused on the substrate at 2.5 mm from the droplet. When the laser light is switched on, the droplet starts to move, and it approaches to the illumination region due to the evanescent electric field components generated by the LN:Fe substrate [Figs. 2(b) and 2(c)].

When the droplet arrives to the illuminate region, it is elongated in the vertical direction due to the light induced evanescent electric field [Fig. 2(d)] and if the distance between the air-paraffin interface and the substrate is short enough ($\leq 0.2$ mm), it makes contact with the substrate [Fig. 2(e)]. Then, the droplet wets the crystal surface and gets trapped on it [Figs. 2(f) - 2(h)] very close but in the boundary of the light beam instead on its center. The inset of Fig. 2(h) shows a magnified fluorescence image of the trapped droplet. Once droplets are trapped on the substrate, they cannot further displace on moving the light beam. This is in accordance with Reference [36] where only covering the substrate with a hydrophobic substrate the water droplet can be moved.

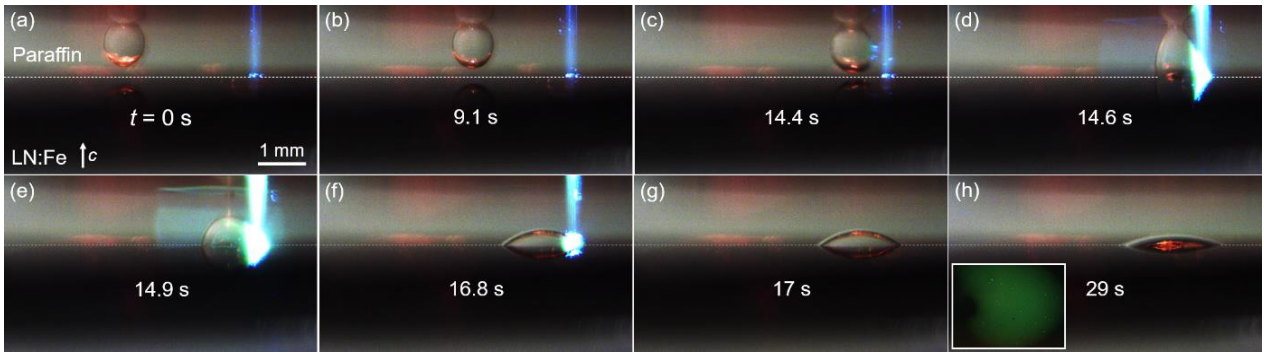

FIG. 2. Photographs taken by the CMOS lateral camera of the droplet manipulation (a)-(d) and trapping (d)-(h) on the z-cut the substrate. The solution contains fluorescein tagged DNA. (Droplet volume $V = 0.2$ μL and laser beam intensity $I = 2.7 \times 10^6$ Wm$^{-2}$). The inset in (h) shows a top view image of the trapped droplet taken with a fluorescence microscope. The time scale is indicated at the top of each image.

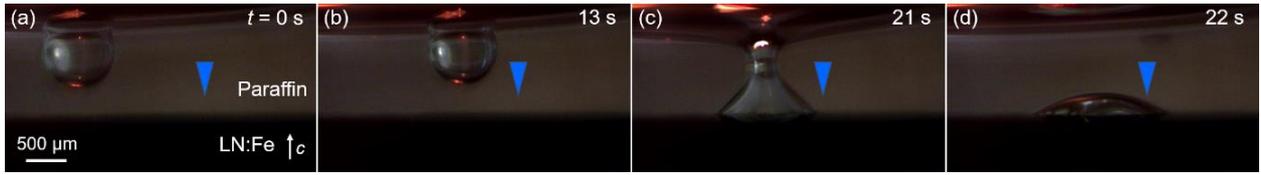

FIG. 3: Photographs taken by the CMOS lateral camera for a deionized water droplet moving after illumination. The position of the previous illumination is indicated by a blue arrow. (a) Addition of the droplet, (b) droplet movement, and (c)-(d) trapping on the crystal. ($V = 0.4$ µL and $I = 2.6 \times 10^7$ Wm$^{-2}$). The time scale is indicated at the top of each image.

In the previous experiments, illumination and manipulation were carried out simultaneously but the manipulation process can be also made after illumination because, as mentioned, the PV fields remain in the dark. This sequential procedure has been mostly used in previous reported work for particle patterning [14] and has been also attempted here for all kind of droplets. The droplets are deposited at the air-paraffin interface after the substrate was illuminated during 1 minute. The droplet moves in the dark to the previously illuminated region and finally traps on the substrate. The process is shown in Fig. 3 for a deionized water droplet.

### 4.2 Droplet trapping at the interface and merging of two droplets

In the previous experiments, the attractive vertical component of the DEP force is high enough to produce droplet trapping on the substrate. However, for certain experiments it can be advantageous to immobilize the droplet at the interface without falling on the PV substrate. This kind of trapping at the interface has been achieved for lower light intensities ($I \sim 7 \times 10^5$ Wm$^{-2}$) and using water and aqueous bio-droplets. After travelling toward the light spot, the hanging droplet is fixed at the position of light illumination at the interface paraffin-air. An illustrative experiment is shown in Fig. 4 where three images of the droplet motion approaching to the light beam (from bottom to top) are presented together with a curve showing the time evolution of the droplet position. This curve clearly shows how the droplet starts to move, accelerates up to a velocity of ~50 µm/s and, finally, stops and keep fixed very close to the light beam but a little shifted from its center. Therefore, PVOT allow two kinds of trapping, either at the interface paraffin-air or on the substrate.

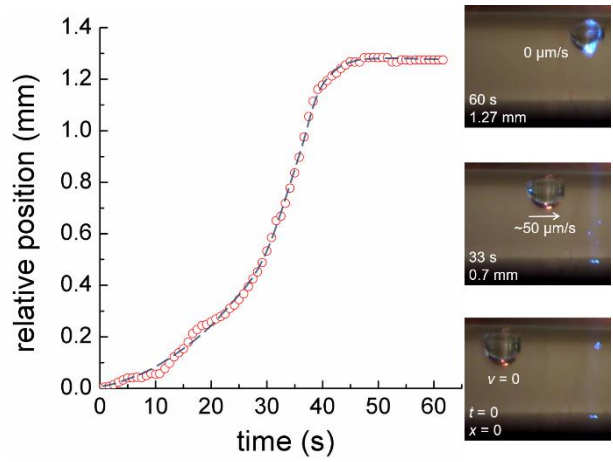

FIG. 4: Time evolution of the water droplet position when moving by the action of PVOT using a *z*-cut LN substrate. Laser peak intensity $7\times10^5$ Wm$^{-2}$. On the right, three insets with photographs corresponding to three different moments of the time evolution. The droplet position, velocity and the time are indicated in each image. Droplet volume *V*=0.2 µL.

In addition, this capability enables another kind of manipulation process with high potential for a variety of applications in optofluidics: to merge two aqueous droplets. The merging experiment obviously requires depositing two droplets at the interface, and it is conducted as follows (Fig. 5). One of the droplets is immobilized by illuminating at its center while the second droplet starts to move towards the light beam [Fig. 5(a) - 5(e)] and finally the two droplets become merged when they get in contact [Fig. 5(f)].

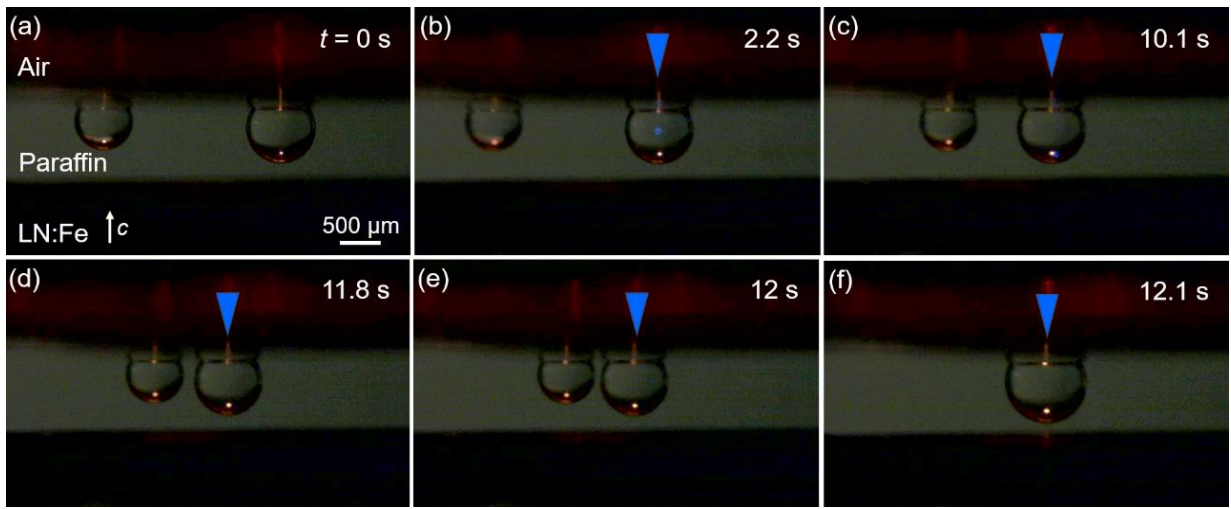

FIG. 5: Photographs taken by the lateral CMOS camera of the merging process using a *z*-cut substrate. The position of the laser beam (not visible) has been marked by a blue arrow. (a) Initial situation. (b) Turning on the laser beam on one droplet. (c)-(e) Shift of the left droplet towards the light beam and so, towards the other droplet. (f) Larger hanging droplet resulting from the merging of the two former ones. Initial droplet volumes are: 0.2 µL and 0.4 µL, respectively. The laser beam intensity is $7\times10^5$ Wm$^{-2}$.

### 4.3 Droplet splitting

Finally, the last kind of experiments have been devoted to attempt droplet splitting. To this end, we have used the parallel configuration (*x*-cut crystals) whose symmetry seems to be more appropriate, because it has a preferential direction (polar axis) on the active surface. Splitting is expected along this direction. In fact, although it was attempted no splitting results were obtained with the isotropic perpendicular configuration (*z*-cut crystals).

In order to get the droplet splitting the substrate has been illuminated ($I = 1.2 \times 10^7$ Wm$^{-2}$) attracting one droplet to the beam spot [see Figs. 6(a) - 6(c)]. Next, by adjusting the beam on the droplet center, it is immobilized [Figs. 6(d) and 6(e)]. Then, the droplet starts elongating along the polar axis [Figs. 6(f) and 6(g)], and, for suitable parameters (in this case, light spot diameter $4\sigma = 160$ µm, droplet diameter $d = 200$ µm and paraffin layer thickness $h = 1$ mm), it splits into two parts of similar volume [Fig. 6(h)].

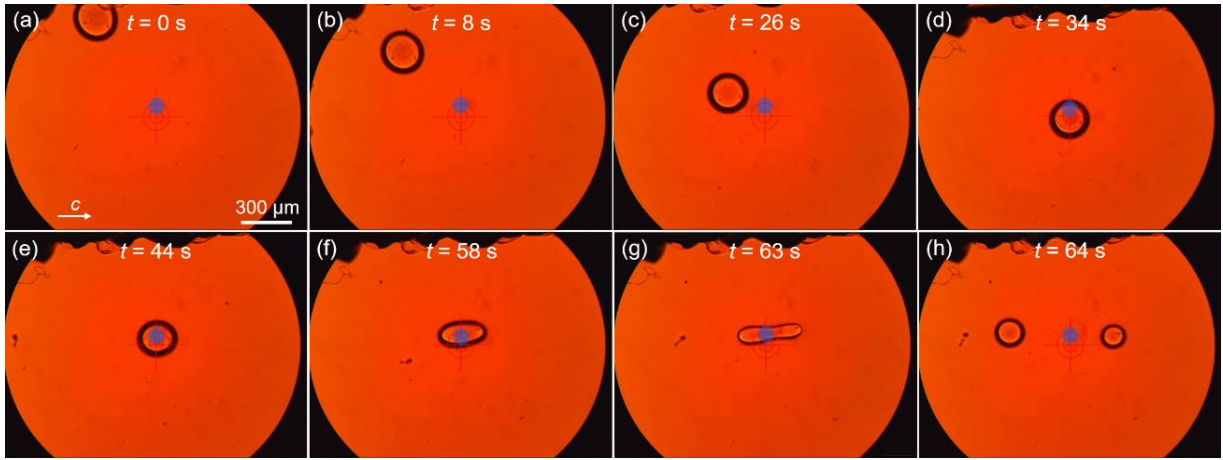

FIG. 6: Sequence of top view photographs showing a water droplet moving toward the light beam (indicated by a blue dot) (a)-(d), and splitting into two droplets (e)-(h), using a *x*-cut LN:Fe crystal. The time scale is indicated at the top of each image. The droplet volume is 0.05 µL and the laser beam intensity is $1.2 \times 10^7$ Wm$^{-2}$.

It is worthwhile remarking that in this experiment, before splitting, the droplet was moved and immobilize at the interface [Figs. 6(a) - 6(c)] demonstrating that these processes are also possible in the parallel configuration of the substrate. In addition, merging has been also obtained for this parallel substrate geometry although the perpendicular configuration could be considered preferable because operation needs a lower light intensity.

## 5. Discussion

In order to make a first qualitative analysis of the experiments, simulations of the PV electric fields and the corresponding DEP force and potential, under Gaussian beam illumination, have been developed. The details of the theoretical model can be found in several previous papers [41–43] and the corresponding simulations have been performed using the software COMSOL Multiphysics. In Fig. 7 the structure of the PV field under Gaussian beam illumination for the two crystal configurations used in the experiments, $z$-cut [Figs. 7(a) - 7(c)] and $x$-cut [Figs. 7(d) - 7(f)] are shown. Schematics of the two geometries are depicted in Figs. 7(a) and 7(d). They show the light beam and the LN:Fe substrate with the direction of the polar axis, and indicate the two planes, parallel and normal to the active surface for which the results of the simulation are shown. Figs. 7(b) and 7(c) ($z$-cut) and Figs. 7(e) and 7(f) ($x$-cut) show the simulations for the electric field vectors (arrows whose length is plotted in a logarithmic scale) and the space charge density generated by the photovoltaic current (using a color code). On the one hand, in $z$-cut, the photovoltaic current (see Eq. 1) generates negative surface charge in the $+c$ face [Figs. 7(b) and 7(c) in red color] and positive surface charge in the opposite $-c$ face (not shown in Fig. 7). On the other hand, in $x$-cut, the redistribution of charge is parallel to the active surface appearing a dipolar charge distribution along the polar axis [Figs. 7(e) and 7(f)]. Looking at planes parallel to the active surface [Fig. 7(c) and 7(f)] one can distinguish the different symmetry of the electric field lines, isotropic for $z$-cut [Fig. 7(c)], and exhibiting a symmetry axis parallel to the polar axis for $x$-cut [Fig. 7(f)]. The different charge and electric field structure of the two configurations is responsible for the differences in the manipulation capabilities (mainly regarding droplet splitting) found for the two substrate configurations.

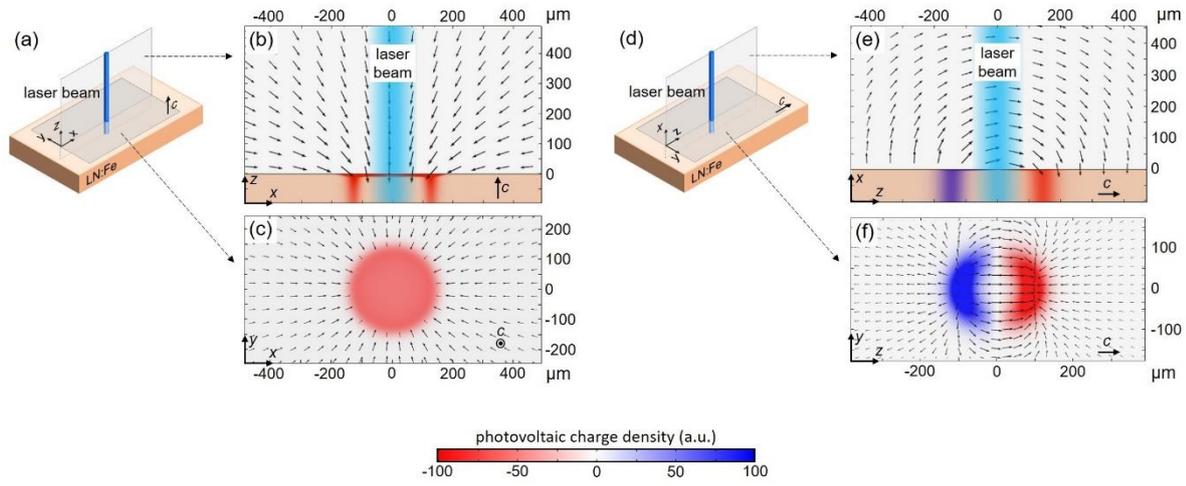

FIG. 7: Spatial distribution of the PV field (arrows) outside the crystal and space charge density (color map) for the perpendicular (a)-(c) and parallel (d)-(e) configurations, along two planes as indicated in the insets. The parallel planes corresponds to a vertical distance of 10 µm from the substrate, i.e. very close to it. The blue laser beam is drawn in (a), (b), (d) and (e). Arrows length, in arbitrary units, is plotted using a logarithmic scale.

Let us focus now our discussion on the experiments of moving, trapping and merging using the perpendicular configuration (*z*-cut) described in sections 4.1 and 4.2. Aqueous droplets are expected to be globally neutral and so, in the presence of PV electric fields, droplets should undergo positive dielectrophoretic forces. In fact, most previous results with PVOT can be qualitatively explained by the action of these DEP forces [14, 15]. In Fig. 8, the spatial distribution of the DEP potential around the light spot is shown using a color map. Moreover, the arrows indicate the DEP forces (the arrows length is plotted using a logarithmic scale). Due to the crystal geometry both, DEP potential and PV fields, have radial symmetry in the XY planes normal to the *z* direction, as it could be seen in the upper inset. Note that the potential well has a ring minimum around the center of the light spot with a radius $r \approx 5\sigma$. When the droplet is relatively far from illumination, the electric field has only a small horizontal component that push the droplet along the interface toward the illuminated region. However, as it approaches the light beam, the vertical component increases and the droplet begins to vertically elongate approaching to the substrate. If the vertical field is strong enough, the droplet makes contact with the substrate and, assisted by the high substrate wettability, become trapped on it. Moreover, the existence of the ring absolute minimum in the DEP potential (clearly shown in the upper inset of Fig. 8) explains the experimental observation of droplet trapping at a position deviated from the center of the light beam (see bottom inset Fig. 8). As clearly seen in these insets the trapped droplet center coincides with the minimum of the calculated DEP potential. The explanation of this subtle experimental feature gives additional support to the theoretical model. If the vertical component of the DEP force close to the light beam is not large enough the hanging droplet immobilizes at the interface, as it was found in Fig. 4. In particular, this is useful for droplet merging: when two droplets are present, both are attracted to the illuminating region by DEP forces and they can be merged as seen in the experiments of Fig. 5. It is remarkable that, the electric field and, so the DEP forces, remain after illumination and so, manipulation can be performed in the dark (Fig. 3) under the action of the same electrical forces invoked when light is present. On the one side, manipulation in the dark is another possibility for this flexible technique that can be advantageous in some cases, e.g. to avoid any damage effect of light on biomaterials. On the other side, it provides a further confirmation that photovoltaic electric fields are the origin of the manipulation processes.

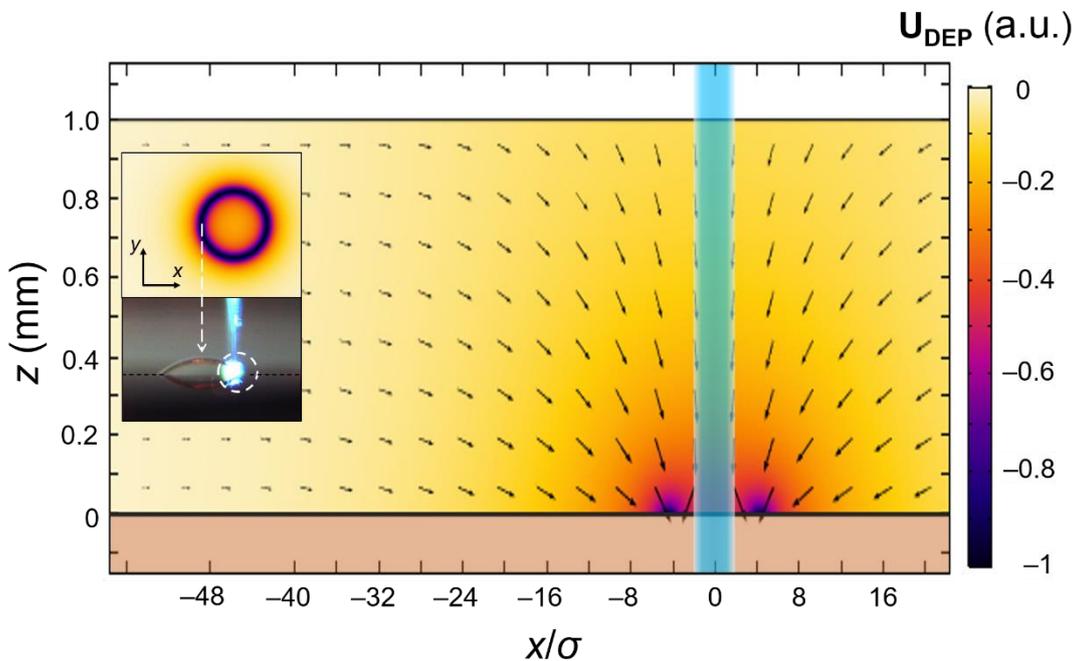

FIG. 8: Color map of the simulated dielectrophoretic potential in $z$-cut under Gaussian beam illumination. Arrows represent the DEP force. Their length, in arbitrary units, is plotted using a logarithmic scale. The upper inset on the left shows a top view, XY plane at 10 µm from the crystal surface, of the DEP potential. The bottom inset shows a droplet trapped on the substrate next to the illuminated beam. The white dashed arrow indicates that the center of the trapped droplet coincides with the position of the minimum DEP potential.

The last experiments consisting in droplet splitting (Fig. 6) have been observed only in the parallel configuration (*x*-cut substrate). Splitting occurs along the polar axis just when the light beam is placed in the center of the droplet. This process can be understood in base of light-induced electric fields simulated in Fig. 7, for *x*-cut. When the droplet is just at the illuminated region, it polarizes along the *c*-axis in such a way that its positive pole orientates toward the negative-charge substrate region and the negative pole toward the positive charge substrate region. Then, electric forces elongate the droplet and they can split it when its size and the distance to the substrate take suitable values. In fact, the time evolution of the light-induced charge distribution on the crystal should favor the process: the distance between the positive and negative substrate charge regions increases with time facilitating the droplet splitting. Finally, other manipulation processes, moving trapping and merging, also observed in this parallel configuration can be qualitatively explained by DEP forces, analogously to the case of the perpendicular geometry.

## 6. Summary and Outlook

A strategy to manipulate water- and bio-droplets by photovoltaic optoelectronic tweezers has been proposed and demonstrated, overcoming a main remaining challenge of these tweezers. Among optoeletronic methods, the technique presented here is simple and cheap. It uses a laser beam that generates the active electric fields in the photovoltaic substrate (LN:Fe) without electrodes or power supplies, and without any further processing of its surface. The method shows key valuable advantages: the droplet hanging at the interface can be manipulated easily and a variety of processes, guiding, trapping at the interface and on the substrate, merging of two droplets and splitting, can be realized. These functionalities stem from the peculiar behavior of the hanging droplet and the flexible manipulation capabilities presented by PVOT, including, those provided by the two crystal geometries (parallel and perpendicular) and, those arising by the operation under illumination or after it in the dark. Therefore, the main outcome of this work is removing the bottleneck existing for flexible manipulation of water and aqueous biodroplets by PVOT providing an efficient tool for optofluidic applications in bio- and nano-technology.

Regarding applications for biotechnology devices, DNA aqueous droplets, diluted sperm and PBS (for cell culture) droplets have been manipulated and so, applications in genetics or cell biology are envisaged. Moreover, one can expect that the manipulation capabilities also extend for any other aqueous droplets with application in diverse areas of biology, physics or chemistry. In addition, the method is well adapted to be implemented in optofluidic devices. For instance, the ability of optically controlled droplet merging is a key capability for applications in chemical synthesis or biotechnological essays [29].

In fact, in some previous works using PVOT, the action on particles immersed in non-polar liquids, has been developed inside microfluidic chips [12, 44]. Furthermore, ferroelectric platforms based in lithium niobate have been already used to generate tiny droplets using the pyroelectric effect [45, 46] or the photovoltaic effect [47]. Then, particularly in the latter case, one could combine in the same optically controlled platform, droplet generation and manipulation functionalities. Further developments in these directions should provide key advances in the near future.

**Acknowledgements**

Financial support from Ministerio de Ciencia, Innovación y Universidades of Spain (MAT2017-83951-R) is gratefully acknowledged. A. Puerto acknowledges the funding under Iniciativa de Empleo Juvenil y Fondo Social Europeo (PEJ2018-003989). Authors also thank Dra. Carmen López-Fernández and Prof. José Luis Bella (Departamento de Biología, Universidad Autónoma de Madrid) for providing the biological material.